\documentstyle[12pt]{article}
\begin{document}
\tolerance=5000
\def\be{\begin{equation}}
\def\ee{\end{equation}}
\def\bea{\begin{eqnarray}}
\def\eea{\end{eqnarray}}
\def\nn{\nonumber \\}
\def\cF{{\cal F}}
\def\det{{\rm det\,}}
\def\Tr{{\rm Tr\,}}
\def\e{{\rm e}}
\def\etal{{\it et al.}}
\def\erp2{{\rm e}^{2\rho}}
\def\erm2{{\rm e}^{-2\rho}}
\def\er4{{\rm e}^{4\rho}}
\def\etal{{\it et al.}}

\  \hfill 
\begin{minipage}{3.5cm}
August 2000 \\
\end{minipage}

\vfill

\begin{center}
{\large\bf 
 AdS/CFT correspondence in cosmology
}

\vfill

{\sc Shin'ichi NOJIRI}\footnote{email: nojiri@cc.nda.ac.jp} 
and {\sc Sergei D. ODINTSOV}$^{\spadesuit}$\footnote{
email: odintsov@ifug5.ugto.mx, odintsov@itp.uni-leipzig.de},

\vfill

{\sl Department of Applied Physics \\
National Defence Academy, 
Hashirimizu Yokosuka 239, JAPAN}

\vfill

{\sl $\spadesuit$ 
Tomsk State Pedagogical University,
634041 Tomsk,RUSSIA
and
Instituto de Fisica de la Universidad de 
Guanajuato \\
Apdo.Postal E-143, 37150 Leon, Gto., MEXICO} 

\vfill

{\bf ABSTRACT}

\end{center}

The attempt to understand if AdS/CFT correspondence may be 
realized as the one between some AdS-like cosmological space 
and CFT living on the boundary is made.
In order to obtain such cosmology we exchange the time and radial coordinates 
in d5 Schwarzschild-anti de Sitter (S-AdS) BH (with corresponding signature 
change). The test on proportionality of free energies from such d5
cosmological space
(after AdS/CFT identification of parameters) and from ${\cal N}=4$ $SU(N)$ 
super Yang-Mills quantum theory is successfully passed.

\newpage

AdS/CFT correspondence \cite{AdS} in its simplest version may be 
understood as duality between 5d gravity and quantum CFT living on 
its boundary (for a review, see \cite{review}). One remarkable 
feature of this duality is that calculations in classical gravity 
on d5 AdS background may provide the answers for dual quantum CFT. 
Very beatiful example of this sort is given by calculation of 
free energy for d5 S-AdS BH \cite{GKP,GKT} which turns out to 
be almost equal to free energy of maximally SUSY Yang-Mills theory 
in the leading approximation on large $N$. The natural question is: 
can we extend above test (or, more generally whole AdS/CFT 
correspondence) to cosmology and how it could be realized on 
practice? In the present note we try to answer this question 
working with specific AdS-like cosmological model obtained by 
exchange of radial and time coordinates in S-AdS BH. Calculating 
analog of free energy for such cosmological model one can show 
that it gives (almost) free energyof maximally SUSY Yang-Mills 
theory but with another mismatch factor than in S-AdS BH case.

We start with the following action of 5-dimensional 
gravity with negative cosmological constant 
$\Lambda=-{12 \over L^2}<0$:
\be
\label{i}
S={1 \over \kappa^2}\int d^5x \left(R + {12 \over L^2}\right)\ .
\ee
In the AdS/CFT correspondence \cite{AdS}, for above SG dual of ${\cal N}=4$
$SU(N)$ or $U(N)$ 
super Yang-Mills theory, we have the following identification
\be
\label{ib}
{L^3 \over \kappa^2}={2N^2 \over 
\left(4\pi\right)^2}\ .
\ee

Let us present brief review on S-AdS BH solution and its thermodynamic
properties. This will be necessary below in attempting to find the
variant of cosmological AdS/CFT correspondence.
A solution of the equation of motion defined by the action (\ref{i}) 
is Schwarzschild-anti de Sitter space, whose metric is given by
\bea
\label{ii}
ds^2&=&-\e^{2\rho}dt^2 + \e^{-2\rho}dr^2 + r^2 \sum_{i,j=1}^3
\hat g_{ij}dx^i dx^j \nn 
\e^{2\rho}&=&{1 \over r^2}\left(-\mu + {k^2 r^2 \over 2} 
+ {r^4 \over L^2}\right)\ .
\eea
Here $\hat g_{ij}$ is the metric of 3 dimensional Einstein 
manifold M$_{\rm E}$, which is defined by
\be
\label{iii}
\hat R_{ij}=k \hat g_{ij}\ .
\ee
Here $\hat R_{ij}$ is the Ricci tensor constructed on $\hat g_{ij}$ 
and $k$ is a constant. In (\ref{ii}), $\mu$ is the parameter 
corresponding to the mass of black hole. If we define the 
parameters $r_h$ and $r_b$ by,
\bea
\label{iv}
r_h^2&=&{L^2 \over 2}\left(-{k \over 2} + \sqrt{
{k^2 \over 4} + {4\mu \over L^2}}\right) \nn
r_b^2&=&{L^2 \over 2}\left({k \over 2} + \sqrt{
{k^2 \over 4} + {4\mu \over L^2}}\right) \ .
\eea
$\e^{2\rho}$ in (\ref{ii}) can be rewritten as
\be
\label{v}
\e^{2\rho}={1 \over L^2 r^2}\left(r^2 - r_h^2\right)
\left(r^2 + r_b^2\right)\ .
\ee
We should note $r_h^2$, $r_b^2>0$. The horizon is given 
by $r=r_h$. The Hawking temperature $T$ is given by
\be
\label{vi}
T={r_h^2 + r_b^2 \over 2\pi L^2 r_h}
={2r_h^2 + {kL^2 \over 2} \over 2\pi L^2 r_h}\ .
\ee

The free energy $F$ would be evaluated by substituting the 
metric given in (\ref{vii}) into the action (\ref{i}):
\be
\label{viii}
F=TS\ 
\ee
after Wick-rotation $t\rightarrow it$. 
In order to avoid the conical singularity, Wick-rotated 
$t$ should be regarded as an angle variable and has a 
period of ${1 \over T}$. 
Using the Einstein equation
\be
\label{ix}
R_{\mu\nu} - {1 \over 2}Rg_{\mu\nu}={6 \over L^2}g_{\mu\nu}\ .
\ee
one gets 
\be
\label{x}
R=-{20 \over L^2}\ .
\ee
Substituting (\ref{x}) into the action corresponding to 
(\ref{i}), one finds the action diverges. Therefore we need 
to regularize the action by introducing 
a cutoff parameter $r_{\rm max}$:
\be
\label{xvb}
S= {V_3L^2 \over \kappa^2 T}\int_{r_h}^{r_{\rm max}}r^3 
\left(-{8 \over L^2}\right)\ .
\ee
Here $V_3$ is the volume of the 3 dimensional 
Einstein manifold M$_{\rm E}$:
\be
\label{xii}
V_3\equiv \int d^3 x \sqrt{\hat g}\ ,
\ee
and ${1 \over T}$ appears due to the periodicity of $t$. 
Subtracting the contribution $S_0$ of the background 
(AdS$_5$)
\be
\label{xvi}
S_0= {V_3L^2 \over \kappa^2 T}
\sqrt{\left(r_{\rm max}^2 - r_h^2\right)
\left(r_{\rm max}^2 + r_b^2\right) \over 
r_{\rm max}^2 \left(r_{\rm max}^2 + r_b^2- r_h^2\right)}
\int_0^{r_{\rm max}}r^3 
\left(-{8 \over L^2}\right)\ ,
\ee
one obtains the following expression of the free energy $F$
\bea
\label{xvii}
F&=&\lim_{r_{\rm max} \rightarrow +\infty} T
\left(S-S_0\right) \nn
&=& -{V_3 \over L^2\kappa^2}\left(r_h^4
 - {kL^2 \over 2}r_h^2 \right) \nn
&=& -{V_3 \over L^2 \kappa^2}\left({1 \over 16}
\left\{
{\pi L^2 T} + \sqrt{\left(\pi L^2 T\right)^2 - kL^2}
\right\}^4 \right. \nn
&& \left.  - {kL^2 \over 8}\left\{
{\pi L^2 T} + \sqrt{\left(\pi L^2 T\right)^2 - kL^2}\right\}^2 
\right)\ .
\eea
Here Eq.(\ref{vi}) is solved with respect to $r_h$: 
\be
\label{xiii}
r_h={1 \over 2}\left\{
{\pi L^2 T}\pm \sqrt{\left(\pi L^2 T\right)^2 - kL^2}\right\}\ .
\ee
In order to reproduce the standard result when $k=0$, 
\be
\label{k0F}
F_{k=0}=-{V_3 \over L^2 \kappa^2}\left(\pi L^2 T\right)^4\ ,
\ee
the sign $\pm$ in (\ref{xiii}) should be $+$. 
Using AdS/CFT table (\ref{ib}), the free energy $F$ when $k=0$ can be 
rewritten in the following form by putting $L=1$:
\be
\label{k0F2}
F_{k=0}=-{\pi^2V_3N^2T^4 \over 8 }\ ,
\ee
which is different from the field theoretical result by a 
factor ${4 \over 3}$ as it was shown in refs.\cite{GKP,GKT}
in all detail
\be
\label{k0F3}
F=-{\pi^2 V_3 N^2 T^4 \over 6}\ .
\ee
The numerical difference may be regarded as due to only leading 
approximation use.

Our next purpose is the attempt to clarify the role of AdS/CFT correspondence 
in the situation when cosmological AdS-like space is considered.
For that one can naively employ the technique which is very similar to
above construction.
Inside the horizon $r<r_h$, if we rename $r$ as $t$ and $t$ 
as $r$ we obtain a metric of a kind of the cosmological model:
\be
\label{vii}
ds^2= - {L^2 t^2 \over \left( r_h^2 - t^2 \right)
\left(r_b^2 + t^2 \right)} dt^2 
+{\left( r_h^2 - t^2 \right)
\left(r_b^2 + t^2 \right) \over L^2 t^2}dr^2 
+ t^2 \sum_{i,j=1}^3 \hat g_{ij}dx^i dx^j \ .
\ee
Thus, we just exchanged the physical role of time and radial 
coordinates \cite{Ryan}.
Since we have only exchanged the coordinates 
$r$ and $t$, the metric, of course, satisfy the Einstein 
equations. Since $t=0$ corresponds to the singularity of 
the black hole, there is a curvature singularity, where the 
square of the Riemann tensor diverges as
\be
\label{Ri}
R_{\mu\nu\rho\sigma}R^{\mu\nu\rho\sigma}\sim {72 \over t^8}\ .
\ee
The singularity at $t=0$ might be regarded as big bang. 
The topology of the space is R$_1\times$M$_{\rm E}$, where 
R$_1$ corresponds to $r$. The metric of R$_1$ vanishes at the 
horizon $t=r_h$. From (\ref{x}), we also find that the scalar 
curvature $R$ is a negative constant, as in AdS.
Note that suggested way is quite limited as narrow class of cosmological
 models may have AdS-like BH cousins.

One can consider an analogue $\tilde F$ of the free energy 
in the metric (\ref{vii}). 
$\tilde F$ would be evaluated by substituting the 
metric given in (\ref{vii}) into the action (\ref{i}) as in 
(\ref{viii}) after ``Wick-rotation'' $r\rightarrow ir$. 
Since $r$ is the time coordinate $t$ in the black hole 
metric (\ref{ii}), $r$ could have a period of ${1 \over T}$ 
after the ``Wick-rotation''. 
Using (\ref{x}), one finds
\bea
\label{xi}
S &=& {1 \over \kappa^2}\int d^5 x\left(-{8 \over L^2}\right) \nn
&=& {V_3 \over \kappa^2 T}\int_0^{r_h} dt 
t^3\left(-{8 \over L^2}\right) \nn
&=& -{2V_3 r_h^4 \over L^2\kappa^2 T}\ .
\eea
Here (\ref{xiii}) is used. 
Then we find the following 
expression of the free energy $\tilde F$:
\be
\label{xiv}
\tilde F=TS= -{V_3 \over 8L^2\kappa^2 }\left\{
{\pi L^2 T} + \sqrt{\left(\pi L^2 T\right)^2 - kL^2}\right\}^4\ .
\ee
Especially when $k=0$, one obtains
\be
\label{xv}
\tilde F_{k=0}= - {2V_3L^6 \left(\pi T\right)^4 \over \kappa^2 }\ .
\ee
Using $\tilde F$ in (\ref{xiv}), 
we obtain the expressions of an analogue $\tilde {\cal S}$ 
of the entropy and that of the 
energy $\tilde E$ as
\bea
\label{xviii}
\tilde {\cal S}&=&- {d\tilde F \over dT} \nn
&=&{\pi V_3 \over 2\kappa^2 }\left\{
{\pi L^2 T} + \sqrt{\left(\pi L^2 T\right)^2 - kL^2}\right\}^3
\left(1 + {\pi L^2 T \over \sqrt{\left(\pi L^2 T\right)^2 
- kL^2}}\right) \\
\label{xix}
\tilde E&=& \tilde F + T\tilde{\cal S} \nn
&=&{V_3 \over 8L^2\kappa^2 }\left\{
{\pi L^2 T} + \sqrt{\left(\pi L^2 T\right)^2 - kL^2}\right\}^3 \nn
&& \times \left(3\pi L^2 T + {3\left(\pi L^2 T\right)^2 + kL^2 
\over \sqrt{\left(\pi L^2 T\right)^2 
- kL^2}}\right)\ .
\eea
Especially when $k=0$, the entropy ${\cal S}$ is given by
\be
\label{k0S}
{\cal S}_{k=0}= {8\pi V_3L^6 \left(\pi T\right)^3 \over \kappa^2 }\ .
\ee
Since the radius of the horizon is not different from the black hole 
case in (\ref{xiii}), we have, when $k=0$, 
\be
\label{k0rh}
r_h=\pi L^2 T\ .
\ee
Then the area of the black hole horizon $A_{\rm BH}$ is given by
\be
\label{k0A}
A_{\rm BH}=V_3(\pi L^2 T)^3\ .
\ee
Putting $\kappa^2=16\pi$ and combining (\ref{k0S}) and 
(\ref{k0rh}), we find
\be
\label{Sbe} 
{\cal S}={A_{\rm BH} \over 2}\ ,
\ee
which seems to be different from the usual result in the black hole 
physics ${\cal S}={A_{\rm BH} \over 4}$. 
It is well-known that definition of entropy in cosmological 
spacetimes is delicated problem (for a recent work on its definition see 
\cite{bousso,HMS}). 
In fact, the cosmological model given here has two horizons at the 
same time. Then the total area $A_{\rm cosm}$ is twice of the area 
of the black hole horizon: 
\be
\label{cosmA}
A_{\rm cosm}=2A_{\rm BH}\ .
\ee
Then from (\ref{Sbe}) and (\ref{cosmA}), we find
\be
\label{cosmS}
S={A_{\rm cosm} \over 4}\ ,
\ee
which does not violate Bousso 
bound \cite{HMS,bousso} ${\cal S}\leq{A \over 4}$.
In order to see that the cosmological model given here has 
two horizons at the same time, one changes the coordinate by
\be
\label{ttau}
\tau={L^{3 \over 2} \over 4\mu^{1 \over 4}}\left\{
\ln\left({1 + {t \over \mu^{1 \over 4}L^{1 \over 2}} 
\over 1 - {t \over \mu^{1 \over 4}L^{1 \over 2}}} \right)
+ \arctan \left({2L^{1 \over 2}\mu^{1 \over 4} \over t}\right)
\right\}
\ee
for $k=0$ case. Then the metric in (\ref{vii}) has the following 
form
\be
\label{confm}
ds^2={1 \over t(\tau)^2}\left(\mu - {t(\tau)^4 \over L^2}
\right)\left(-d\tau^2 + dr^2\right) + t(\tau)^2\sum_{i=1}^3 
\left(dx^i\right)^2\ .
\ee
Here $t$ in (\ref{confm}) is given by solving (\ref{ttau}). 
Near the horizon $t\sim \mu^{1 \over 4}L^{1 \over 2}$, we have
\bea
\label{conf1}
1-{t \over \mu^{1 \over 4}L^{1 \over 2}}&\sim& \e^{-{4
\mu^{1 \over 4} \over L^{3 \over 2}}\tau} \\
\label{conf2}
ds^2&\sim&{4\mu^{1 \over 2} \over L}\e^{-{4
\mu^{1 \over 4} \over L^{3 \over 2}}\tau} \left(
-d\tau^2 + dr^2\right) + \mu^{1 \over 2}L\sum_{i=1}^3 
\left(dx^i\right)^2 \nn
&=& -{4\mu^{1 \over 2} \over L}\e^{-{2
\mu^{1 \over 4} \over L^{3 \over 2}}\left(x^+ + x^-\right)} 
dx^+dx^- + \mu^{1 \over 2}L \sum_{i=1}^3 \left(dx^i\right)^2 \ .
\eea
Here
\be
\label{conf3}
x^\pm\equiv \tau \pm r\ .
\ee
Eq.(\ref{conf2}) tells that the cosmological model given here 
has two horizons corresponding to $x^+\rightarrow +\infty$ and 
$x^-\rightarrow +\infty$ at the same time $t=
\mu^{1 \over 4}L^{1 \over 2}$.
Hence, we demonstrated that our proposal being limited to narrow class 
of cosmological spacetimes gives correct value for cosmological entropy.

Taking (\ref{ib}) and putting $L=1$, Eq(\ref{xiv}) can be 
rewritten as 
\be
\label{xivb}
\tilde F= -{V_3N^2 \over 4(4\pi)^2 }\left\{
{\pi T} + \sqrt{\left(\pi T\right)^2 - k}\right\}^4\ .
\ee
Especially when $k=0$, one finds 
\be
\label{xivc}
\tilde F_{k=0}= -{V_3N^2\pi^2T^4 \over 4}\ ,
\ee
which is twice of the free energy (\ref{k0F2}) of S-AdS black hole. 
Thus, our example demonstrates that could be cosmological AdS/CFT
correspondence
between some cosmological (AdS-like) space and ${\cal N}=4$ $SU(N)$ 
super Yang-Mills theory
 living on its boundary \footnote{ The original S-AdS BH we started with 
is dual to ${\cal N}=4$ $SU(N)$ super Yang-Mills theory which lives on the 
boundary of S-AdS BH. Having no proof one
can conjecture quite naturally that after exchange the role of time and 
radius the obtained cosmological AdS-like Universe is still dual to the 
same boundary 
super Yang-Mills theory. The reason is that at fixed time 4d geometry of
5d cosmological spacetime is the same as 4d geometry of S-AdS BH at 
fixed radius. Hence, we dont expect any effect of time-radius exchange 
to dual QFT. Of course, this should be carefully checked.}  
 (at least on the level of free energy). It suggests the way of 
calculation of QFT free energy and entropy starting from 
cosmological background. Note that our explicit example gives 
mismatch factor 3/2 between ``cosmological''  and super Yang-Mills 
free energy. It is interesting that above description may be 
relevant also for construction of entropy in 
cosmology\cite{Gibbons},
which is not easy in case of expanding Universe.

Let us give few more remarks on the spacetime described by the metric 
(\ref{vii}). If we change the coordinate by
\be
\label{t1}
t^2 = {r_h^2 + r_b^2 \over 2}\cos {2\tau \over L}
+ {r_h^2 - r_b^2 \over 2}
= L^2\sqrt{{k^2 \over 4} + {4\mu \over L^2}}\cos {2\tau \over L}
- {L^2k \over 2}\ ,
\ee
one can rewrite the metric (\ref{vii}) in the following form:
\bea
\label{t2}
ds^2 &=& - d\tau^2 + {\left({k^2 \over 4} + {4\mu \over L^2}\right)
\sin^2 {2\tau \over L} \over \sqrt{{k^2 \over 4} 
+ {4\mu \over L^2}}\cos {2\tau \over L}
- {k \over 2}}dr^2 \nn
&& + \left(  L^2\sqrt{{k^2 \over 4} + {4\mu \over L^2}}
\cos {2\tau \over L}- {L^2k \over 2}\right)\sum_{i,j=1}^3 
\hat g_{ij}dx^i dx^j \ .
\eea
In (\ref{t2}), $\sin{2\tau \over L}=0$ corresponds to the 
horizon and $\cos {2\tau \over L} = {k \over 2\sqrt{{k^2 \over 4} 
+ {4\mu \over L^2}}}$ corresponds to the 
curvature singularity. Therefore the lifetime of the universe 
is ${\cal O}(L)$. Usually $\mu$ is not negative since $\mu$ 
corresponds to mass of the black hole. If we consider the case 
of $\mu<0$ and $k<0$, however, there does not appear the 
curvature singularity since ${|k| \over 2\sqrt{{k^2 \over 4} 
+ {4\mu \over L^2}}}>1$. Epecially if one considers the limit 
$\mu\rightarrow - {k^2L^2 \over 16}$ and rescales the coordinate 
$r$ by $r\rightarrow {r \over \sqrt{{k^2 \over 4} 
+ {4\mu \over L^2}}}$, we obtain the following metric:
\be
\label{t3}
ds^2 = - d\tau^2 + {2 \sin^2 {2\tau \over L} \over |k|}dr^2 
+ {L^2|k| \over 2} \sum_{i,j=1}^3 \hat g_{ij}dx^i dx^j \ .
\ee
In the metric (\ref{t3}), the radius of the Einstein manifold 
M$_{\rm E}$, whose metric is expressed by $\tilde g_{ij}$, is 
constant. When we impose a periodic boundary condition on $r$, 
which is not always necessary in the Minkowski signature, $r$ 
expresses the coordinate of S$_1$. Then the radius of S$_1$ 
becomes a periodic function of $\tau$. Hence, one actually has 
two metrics with time topology S$_1$ or R$_1$.    

In the metric (\ref{t3}), since $\mu= - {k^2L^2 \over 16}$, 
from (\ref{iv}), we find 
\be
\label{AdSNariai}
r_h^2=-r_b^2={L^2 |k| \over 4}\ .
\ee 
When $r_b^2<0$, the radius $\sqrt{-r_b^2}$ corresponds to 
the inner horizon. Then the metric (\ref{t3}) corresponds to 
a limit where two horizons coincide with each other, as in 
the Nariai limit in Schwarzschild-de Sitter solution. Using 
(\ref{vi}), one finds the temperature $T$ vanishs:
\be
\label{AdSNT}
T=0\ .
\ee
In spite of (\ref{AdSNT}), from Eq.(\ref{xiv}), we get the 
analogue of the free energy $\tilde F$ is finite:
\be
\label{AdSNF}
\tilde F= -{V_3|k|^2L^2 \over 32\kappa^2 }\ .
\ee
Using (\ref{ib}), the expression in 
(\ref{AdSNF}) can be rewritten as 
\be
\label{AdSNFb}
\tilde F= -{V_3|k|^2N^2 \over 16L\left(4\pi\right)^2 }\ .
\ee
As one sees from here, when 3d space is flat the above expression 
is zero.
It coincides with the expectations: QFT result for Casimir energy 
of super Yang-Mills theory with four supersymmetries on 4d flat space is zero.
On the other side, when 3d space has constant curvature it is not zero.
In this case it should be identified with vacuum energy (Casimir energy) 
(for calculation of Casimir energy in AdS/CFT set-up see 
refs.\cite{casimir}).
Indeed, the calculation of Casimir energy from QFT side gives 
non-zero result when space part of metric represents 3d sphere 
and zero result when it represents 3d hyperbolic space.
However,  the only global characteristic of spacetime (negative curvature) 
has been concerned so the  
above example may correspond to  
Casimir energy in compact 
space H$_3/\Gamma$. The corresponding calculations have been done for 
various isometry groups (see for a review \cite{bytsenko}). 
The general result 
for super Yang-Mills theory (same boundary conditions
 for all fields of supermultiplet are assumed) under consideration
is following:
\be
\label{W}
W={cV_3N^2 \over L(4\pi)^2}
\ee 
where numerical constant $c$ is defined by features of space 
H$_3/\Gamma$. Hence, we again obtain the qualitative agreement 
with the expression (\ref{AdSNFb}).

In summary, using simple examples of cosmological AdS-like space 
obtained from S-AdS BH we speculated on realization 
of AdS/CFT correspondence on ``cosmological'' level. Clearly, more 
work is necessary in order to understand such version of AdS/CFT 
duality (if it can exist as real duality and it is not the 
occasional coincidence). The results of this note do not prove 
the possibility of such cosmological AdS/CFT duality. Rather, 
that can be considered as some hint towards to its existance. 
Nevertheless, it is interesting that similar construction 
is possible also in higher derivative gravity which also 
admits S-AdS BHs \cite{NO} 
permitting to check the cosmological AdS/CFT correspondence even in
next-to-leading order. This will be done elsewhere. 

\ 

\noindent
{\bf Acknoweledgements.} 
We wish to thank O. Obregon for very helpful discussion.
The research by SDO has been supported in part by 
CONACyT (CP,ref.990356
and grant 28454E) and in part by NORDITA.

\end{document}